\documentstyle[preprint,prb,aps]{revtex}
\draft
\begin{document}
\title{Nonlinear acoustic and microwave absorption in glasses}
\author{M. Kirkengen$^{(1)}$ and Yu. M. Galperin$^{(1,2})$}
\address{ $^{(1)}$Department of Physics, University of Oslo,  P. O. Box 1048  
Blindern, N 0316 Oslo, Norway \\ $^{(2)}$A. F. Ioffe  
Physico-Technical  Institute, 194021 St. Petersburg, Russia}  
\date{\today} \maketitle

\begin{abstract} 
A theory of weakly-nonlinear low-temperature relaxational absorption of
acoustic and electromagnetic waves in dielectric and metallic glasses
is developed. Basing upon the model of two-level tunneling systems we
show that the nonlinear contribution to the
absorption can be anomalously large. This is the case at low enough
frequencies, $\omega \tau_0 (T) \ll 1$, where $\tau_0 (T)$ is the minimal
relaxation time for two-level systems with the inter-level splitting
$\sim k_BT$.  In dielectric glasses, the lowest-order nonlinear contribution is
proportional to the wave's intensity. It is negative and exhibits 
anomalous frequency and temperature dependencies, $\Delta \Gamma
/\Gamma_0 \propto [\omega \tau_0 (T)]^{-1/2}T^{-2}$. In metallic
glasses, the nonlinear contribution is also negative, and it is
proportional to the {\em square root} of the wave's
intensity and to the frequency. Numerical estimates show that the
predicted nonlinear 
contribution can be measured experimentally.

\end{abstract}

\narrowtext

\maketitle

\section{Introduction}
 
Acoustic and dielectric properties of glasses were intensively studied
during the last decades.  The most remarkable result\cite{ZP} of these
studies was the discovery of universal low-temperature properties that
are only weakly dependent on the chemical composition of the
glass. These ``anomalous'' specific properties include low-temperature
specific heat, thermal conductivity, thermal expansion, propagation of
ultrasound, dielectric loss, electric and acoustic echo and some other
properties. An important step towards an understanding of the
``anomalous'' properties of glasses was the introduction of the
tunneling model\cite{AHV,Phil}. According to this model, there exist
{\em two-level systems} (TLS) associated with local tunneling states in
double-well potentials. These states are characterized by an energy
difference $2\Delta$ between the minima and a tunnel splitting
$2\Lambda$, the energy spacing being 
\begin{equation} \label{Eo}
2 \epsilon \equiv
2\sqrt{\Delta^2 + \Lambda^2}\,.
\end{equation} 
 The parameters $\Delta$ and $\Lambda$
are random, the 
distribution of $\Delta$ and $\ln \Lambda$ being assumed constant.
 There are several review
articles\cite{HA,HR,Phil1,Phil2,Black} where experimental data and
their interpretation on the basis of the tunneling model are given.   

Two different absorption mechanisms are usually discussed in
connection with the TLS. The first one is resonant absorption, which
is a direct absorption of acoustic 
(or microwave) quanta $\hbar \omega$ accompanied by transitions
between the levels of TLS. Resonant absorption of an acoustic wave can
be expressed as\cite{HA} 
\begin{equation} \label{gr1}
 \Gamma^{(\text{res})} =\alpha (\omega/s)\, \tanh \left(
\hbar \omega / 2 k_B T\right)
\end{equation}
where $\alpha$ is a dimensionless coupling constant, $s$ is the sound
velocity, while $T$ is the temperature.  The case of microwave differs
from Eq.~(\ref{gr1}) only by the coupling constant, so we will discuss
below only the case of acoustic waves.
At high enough temperature ($\hbar \omega \ll k_B T$) Eq.~(\ref{gr1}) yields
\begin{equation} \label{gr2}
 \Gamma^{(\text{res})} \approx \alpha \hbar\omega^2/2sk_BT \, .
\end{equation}
Another mechanism - {\em relaxational absorption} - is due to
modulation of the populations of the TLS levels by the alternating
deformation field created by the acoustic wave. This modulation is due
to the time-periodic variation in the inter-level spacing $2 \epsilon$. 
This variation produces, in its turn, a modulation  of
the levels' populations  
 which lags in phase the variation of $\epsilon$. This lag leads
to the energy dissipation. The {\em linear} in the acoustic intensity
relaxational 
absorption can be estimated as\cite{HA,Jackle}
\begin{equation} \label{rel1}
\Gamma^{\text{(rel)}} \approx \frac{\alpha}{s}\left\{ \begin{array}{ll}
\tau^{-1},& \omega \tau_0 \gg 1; \\
\omega, & \omega \tau_0 \ll 1, \end{array} \right.
\end{equation}
where $\tau_0 (T)$ is the minimum relaxation time of the TLS having
$\epsilon \approx k_BT$. Comparing Eq.~(\ref{gr1}) with
Eq.~(\ref{rel1}) we conclude that the relaxational absorption always
predominates at $\omega \tau_0 \ll 1$. If $\omega \tau_0 \gg 1$ the ratio
$\Gamma^{\text{(res)}}/\Gamma^{\text{(rel)}}\approx
\hbar \omega^2 \tau_0/k_BT$
can be either greater or less than one under experimentally
accessible conditions.

{\em Nonlinear} absorption was mostly studied for the resonant
mechanism. Actually, the 
nonlinear resonant absorption of sound is one of the basic effects
confirming the existence of TLS in disordered materials (see e. g. 
Ref.~\onlinecite{HA}).  The factor $\tanh (\hbar
\omega/2k_B T)$ in Eq.~(\ref{gr1}) has a transparent physical meaning
-- it is the 
difference in equilibrium populations of the lower and upper levels of
a {\em resonant} TLS having the inter-level spacing $\hbar \omega$.  The
reason for the nonlinearity is that intense sound equalizes the levels'
populations, which again results in a decrease in the
absorption. Experimentally, the effect takes place at very
low acoustic intensities\cite{Jackle}. Consequently, the resonant
absorption decreases with the intensity increase, and at large enough
amplitudes only the relaxational contribution can be observed. As a
result, the {\em nonlinear} relaxational absorption becomes important.

The source of the  nonlinear relaxational absorption is a strong
modulation of the inter-level spacing of the relevant TLS (with the
inter-level distance $\approx k_B T$) by the sound wave. As a result,
the relevant TLS are able to absorb energy only during a part of the
sound period. Consequently, the total absorption decreases with the
sound amplitude. A theory for the nonlinear relaxation absorption in
insulation glasses has been developed in\cite{Gal1,Laikh}, the authors
were concentrated on the case of large acoustic intensities. It seems
that it is difficult to realize such a regime in a realistic
experimental situation with insulating glasses. However,
the nonlinear relaxational absorption was observed experimentally in metallic
glasses (PdSiCu, PdSi, PdNiP)\cite{MG} as a specific two-stage
nonlinear behavior (successive decrease of resonant and then
relaxational absorption as the intensity increases). The relevant
theory for the case of large intensities  has been developed in
Refs.~\onlinecite{GGP}.  

In this paper we show that one can
expect pronounced effects in nonlinear relaxational absorption even at
relatively low intensities. Namely, we claim that at low enough acoustic
frequencies  the nonlinear corrections to the linear absorption
coefficient $\Gamma^{\text{(rel)}}_0$ are rather pronounced and
possess unusual frequency and temperature dependencies. We hope that
the regime of ``weak nonlinearity'' is easier accessible for the
experiments, especially  with insulating glasses.  

According to the theory for the strongly nonlinear
regime\cite{Gal1,Laikh}, significant deviation from the linear
relaxational absorption takes place at $d \gtrsim k_BT$ where $d$ is
the amplitude of modulation of the inter-level spacing ($\sim k_BT$)
of the relevant TLS.  Consequently, one can think
that small nonlinear corrections to the linear absorption coefficient
behave as $(d /k_BT)^2$. However, the energy difference $2 \epsilon$
between the levels depends on both the diagonal splitting $\Delta$ and
the tunnel coupling $\Lambda$ according to Eq.~(\ref{Eo}).  Furthermore, the
relaxation time $\tau$ for a given TLS is also dependent on its
parameters $\Delta$ and $\Lambda$. This dependence can be expressed as
(cf with Refs.~\onlinecite{HA,Black}) \begin{equation} \label{t0}
\frac{1}{\tau (\Delta, \Lambda)} = \frac{1}{\tau_0
(T)}\left(\frac{\epsilon}{k_BT}\right)^{\beta+1}
\left(\frac{\Lambda}{\epsilon}\right)^2 \coth
\left(\frac{\epsilon}{k_BT}\right) \, .  \end{equation} The first
factor has a meaning of a {\em minimal} relaxation time for the
systems with the inter-level spacing $\epsilon \approx k_B T$. The
fact that the inter-level transitions can take place only if the wells
are coupled is described by the factor $(\Lambda/\epsilon)^2$. This
factor is maximal at $\Delta =0$ or at $\Lambda = \epsilon$. In the
factor $(\epsilon/k_BT)^{\beta +1}$ one power of $\epsilon$ is due to
deformational coupling between TLS and internal degrees of
freedom\cite{HA,Black}, while $\epsilon^\beta$ describes the energy
dependence of the density of states for the internal degrees of
freedom. In insulating glasses 
these are phonons ($\beta=2$) while in metallic glasses the relaxation
is due to electronic excitations near the Fermi level ($\beta
=0$)\cite{Black}. The last factor represents the sum of the occupation
numbers for 
the excitation responsible for the absorption and emission
processes. Indeed, $\coth (\epsilon/k_BT) =2N(2\epsilon) +1$ where $N$
is the Planck 
function.  Consequently, nonlinear effects in absorption are
determined by the ratio between the amplitude of modulation of the
diagonal splitting, $d$, and the characteristic energy, $\delta$, which
is a combination of the parameters $\Delta$ and $\Lambda$. This
combination is determined by the dimensionless product $\omega
\tau_0$, and it appears that $\delta \ll k_BT$ at $\omega \tau \ll 1$
(see below).
As a result, at low frequencies nonlinear effects must be
pronounced. They must also show specific frequency dependence. Having
in mind that at $\omega \tau \ll 1 \quad$ $\Gamma_0^{\text{(rel)}} \propto
\omega$ and temperature-independent, one can expect that nonlinear
effects can be detected also from anomalous temperature and frequency
dependencies of absorption. This is the main message of the paper.

Below we will sketch the basic theoretical
approach to calculate both linear and weakly nonlinear
absorption. Then we will concentrate 
on the nonlinear 
correction to the  absorption coefficient for relatively low
amplitudes, both for insulating and metallic glasses.

\section{Theoretical basis}

The Hamiltonian of a TLS in the external ac field  can be written
as\cite{Jackle}   
\begin{equation}
{\cal H}=(\Delta + d\, \cos \omega t) \sigma_z -\Lambda \sigma_x\,,
\end{equation}
where $\sigma_{x(z)}$ are Pauli matrices, $\Delta$ is the energy gap
between the levels of the isolated potential wells, while $\Lambda$
is the tunnel matrix element, $d=\gamma_{ik}u_{ik}^{(0)}$ in the case
of sound wave ($\gamma_{ik}$ is the deformational potential of the
TLS, $u_{ik}^{(0)}$ is the amplitude value of the deformation
tensor). In the case of electromagnetic wave $d=\eta {\cal E}_0$,
where $\eta$ is the dipole moment of the TLS while ${\cal E}_0$ is the
amplitude of the electric field. The 
corresponding change in the tunneling transmittance $\Lambda$ we
shall assume to be small and neglect\cite{Jackle,Black}. 

According to the TLS model\cite{AHV,Phil} we assume $\Delta$ and
$\Lambda$ to be independent random quantities, $\Delta$ and $\ln
\Lambda$ being uniformly distributed over a broad range of values in
comparison with the temperature. We shall also assume the
deformational potential $\gamma_{ik}$ to be a random quantity,
uncorrelated with $\Delta$ and $\Lambda$ , the distribution of which
has a maximum  (a similar assumption can be introduced
concerning $\eta$).   
 
Following Refs.~\onlinecite{Gal1,Laikh}, we consider the case of
relatively low frequencies when the energy of the acoustic quantum
$\hbar \omega$ is much less that the characteristic inter-level
distance $\delta$ of the TLS which make a contribution to the
nonlinear absorption (we shall estimate $\delta$ later). At  
$
\hbar \omega \ll \delta
$
one can employ the adiabatic approximation and neglect 
time derivatives of the external field while solving the Schr\"odinger
equation for the TLS. In this approximation the TLS is characterized by
the time-dependent spacing $2 \epsilon (t)$, 
\begin{equation} \label{espacing}
\epsilon (t) = \sqrt{( \Delta + d \cos \omega t
)^2 + \Lambda ^ 2}\, , 
\end{equation}
and the occupation numbers of the upper ($n$) and the lower ($1-n$) levels.
The non-equilibrium occupation numbers can be found from the balance
equation
\begin{equation} \label{be}
\frac{dn}{dt} = -\frac{n-n_0(t)}{\tau (t)}
\end{equation} 
where $\tau (t)$ and $n_0 (t)$ are given by the substitution $\epsilon
(t) \rightarrow \epsilon$ into the Eq.~(\ref{t0}) and into the expression 
\begin{equation} \label{n_0}
n_0 = \left[\exp (2\epsilon/k_BT)+1 \right]^{-1}
\end{equation}
for the equilibrium occupation number. According Eq.~(\ref{be}), the
dynamics of the levels' population is characterized by a
time-dependent relaxation time. The solution of Eq.~(\ref{be}) for a
strongly nonlinear regime ($d \gg k_BT$) has been analyzed in
Refs.~\onlinecite{Gal1,Laikh} for the case of insulating glasses and
in Ref.~\onlinecite{GGP} for metallic glasses. 
Below we shall present a theory for the regime of weak nonlinearity
which seems easier to realize.

\section{Calculations}

In the following we will use the subscript 0 to indicate linear results.
The power absorbed by a single TLS can be determined by the
expression\cite{Gal1} 
\begin{equation} \label{abs}
p(\Delta, \Lambda)  =  \frac{\omega}{\pi}\int_0^{2 \pi/\omega} n\frac{ d
\epsilon}{d t} \, d t 
\end{equation}  
The contributions of individual TLS must be added and such a summation
can be performed in a conventional way using the distribution function
of the random parameters $\Delta$ and $\Lambda$ and replacing the
deformational potential $\gamma_{ik}$ by its average value. The
distribution function is usually\cite{AHV,Phil} chosen as 
$$N(\Delta, \Lambda)= 2N/\Lambda $$
where $N$ is the the density of states per energy interval while
$1/\Lambda = d(\ln \Lambda)/d \Lambda$ describes a smooth distribution
of $\ln \Lambda$. The coefficient 2 is introduced because we use the
notation $2 \epsilon$ for the energy interval.  
 This gives for the total absorption
\begin{equation} \label{totabs}
P = { 2N \int_0^\infty d \Delta \int_0^\infty
\frac{d \Lambda}{\Lambda}\,  p(\Delta,\Lambda,d)}\, .
\end{equation}  

To analyze the absorption one can use the exact periodic in
time solution of Eq.~(\ref{be}) to obtain\cite{Gal1}
\begin{eqnarray} \label{esol}
P &=& \frac{N}{\Theta}
\int_0^\infty
\int_0^\infty \frac{d \Delta \, d \Lambda}{ k_B T\,\Lambda}
\left(1 - e^{ -\int_0^{\Theta}
dt_1/\tau (t_1)}\right)^{-1}
\nonumber \\ &\times &
\int_0^{\Theta}
\int_0^{\Theta} \frac{dt\, dt'\,{\dot \epsilon} (t){\dot
\epsilon}(t-t')}{\cosh^2 [\epsilon (t-t')/k_B T] } 
e^{- \int_0^{t'} dt_1/\tau
(t-t_1)} .
\end{eqnarray}
Here $\Theta =2\pi/\omega$ is the period of the wave.
We will first consider the case of dielectric glasses.
In the linear approximation one has 
$${\dot \epsilon_0} (t) = -(\Delta/\epsilon)\, \omega d \sin \omega t\, ,
\quad \epsilon = \sqrt{\Delta^2 + \Lambda^2}\, , $$
while the relaxation rate is time-independent and given by
Eq.~(\ref{t0}). {}From Eq.~(\ref{esol}) one obtains the well-known
expression for the relaxational absorption\cite{Jackle}
\begin{equation} \label{la}
P=\frac{N \omega }{2k_BT}\int_0^\infty \int_0^\infty \frac{d \Delta\,
d \Lambda}{\Lambda \,\cosh^2 (\epsilon/k_BT)}\, \frac{\omega \tau}{1 + \omega^2
\tau^2} \, .  
\end{equation}  
Since $\tau(\Delta,\Lambda) \propto \Lambda^{-2}$ the typical value of
$\Lambda$ in this integral is $k_B T$ at $\omega \tau_0 \gg 1$ and
$k_BT\sqrt{\omega \tau_0}$ at $\omega \tau_0 \ll 1$. It is the last
regime that produces anomalous nonlinear effects. In the linear case
at $\omega \tau_0 \ll 1$ one obtains
\begin{equation} \label{la1}
P_0=(\pi^2/16)N \omega d^2
\end{equation}

As we shall see, the nonlinear corrections to the absorption 
are of greatest importance for low values of the inter-level spacing 
$\epsilon \ll k_BT$. Consequently, one can still use the asymptotic
expression for $\coth (\epsilon/k_BT) \approx k_B T/\epsilon$ in
Eq.~(\ref{t0}). As a result, in the case of $\beta =2$ (i. e. for 
insulating glasses) 
$\tau =\tau_0 \, (k_BT/\Lambda)^2$,
 and it is {\em time-independent}. That leads to important
simplifications in Eq.~(\ref{esol}). Indeed, to get nonlinear
corrections to the absorption one has to expand only the function 
$${\dot \epsilon} (t) = -\frac{\Delta + d \cos \omega t}{\sqrt{(\Delta
+ d \cos \omega t)^2 + \Lambda^2}} \, \omega d\, \sin \omega t
$$
in powers of $d$ up to second order, and then perform the
integrations over $t,t',\Delta$ and $\Lambda$. Such a calculation can
be easily performed using any computer algebra package (we used
``Maple''). The result can be expressed as
\begin{equation} \label{res1}
(P-P_0)/P_0 = - c_d \left(d/k_BT \right)^2
[\omega \tau_0 (T)]^{-1/2}\, . 
\end{equation}
Here $c_d = (3/32)(4 \sqrt{2}-1) \approx 0.44$ is the numerical
factor.

 In the case of metallic glasses, the linear calculations are performed
analogous to the dielectric case. Making use of the $\Lambda$ dependence of the
relaxation time that emphasizes small  $\Lambda$
we get $P_0 \approx 0.38 N \omega d^2$ for the most
interesting case  $\omega\tau \ll 1$. For the nonlinear
corrections, however, we can no longer assume the transition time to
be time-independent. Indeed, calculations show that the main
contribution to the nonlinear absorption comes from the expansion of the
exponential in Eq.  (\ref{esol}). 
Expanding the power $p$ absorbed by a single TLS up to the
fourth order in the wave's amplitude and subtracting the quadratic
contribution $p_0$ (which is responsible for the linear absorption
coefficient)  we obtain  
$$p-p_0 = \frac{\nu \,\left(\nu \sin \theta -8\right) \pi \sin^2 \theta
\cos^4 \theta}{\left(4+\nu \sin \theta \right)\left(1+\nu \sin \theta
\right)}\, \frac{d^4}{\epsilon^2}\, .$$  
Here we denote $\nu = (\omega \tau_0)^{-1}, \ \Lambda = \epsilon \sin
\theta, \ \Delta = \epsilon \cos \theta$.

As can be seen, this expression diverges strongly for small
$\epsilon$, so it is not possible to obtain an exact quantitative
estimate from our simplified calculation. In order to make an order-of
magnitude-estimate, however,
we may cut off the integral in the lower limit, at $\epsilon = |d|$.
This assumption yields  for $\omega \tau \ll 1$
\begin{equation}
\label{res1a} (P-P_0)/P_0 = - c_m (\omega \tau_0)^{-1}\left(|d|/k_B T \right)  
\end{equation} 
with $c_m$ is of the order 1.
We observe that the nonlinear contribution is proportional to $|d|$,
i. e. to the {\em square root} of the intensity rather than to the
intensity, as in usual cases. 
This dependence arises from the fact that nonlinear effects are
especially important for the TLS with small inter-level splitting
$\epsilon \lesssim |d|$. Indeed, these systems are strongly perturbed
by the acoustic wave. However, the probability to find a TLS with
small splitting is proportional to $|d|$. Furthermore, at small
$\epsilon$ the relaxation time the $\tau_0 (\epsilon)$ increases as
$\epsilon^{-1}$ that leads to a decrease of the contribution of the
low-energy TLS. Collecting of all the factors leads to the estimate   
(\ref{res1a}). Unfortunately, our calculation based on the expansion
in powers of $d/\epsilon$ cannot provide the numerical
factor. However, it yields the proper functional dependencies. 

In the high-frequency limit, $\omega \tau \gg 1$, the nonlinear 
contribution is $\propto (d/k_BT)^2$.

\section{Discussion}
Since $P_0 \propto \omega$ and independent of $T$, the nonlinear
contribution to the sound absorption in the {\em dielectric glasses}
appears $\propto \sqrt{\omega}$, the temperature 
dependence (at fixed $d$) being $\propto T^{-2} [\tau_0 (T)]^{-1/2}
\propto T^{-1/2}$.  

Analyzing the relevant integrals, one comes to the conclusion that
the characteristic inter-level spacing of the TLS responsible for the
absorption is $\delta \sim k_B T (\omega \tau_0)^{1/4} \le k_BT$. This
fact confirms that one can employ low-energy asymptotics in
Eq.~(\ref{t0}) and replace $\cosh^{-2} (\epsilon/k_B T) \rightarrow
1$ in Eq.~(\ref{esol}). Inequality $\hbar \omega \ll \delta$ appears
met for realistic parameters of the glass. 

Now, let us give a rough estimate of the intensities needed to obtain
a significant nonlinear contribution for dielectric glasses. One can
estimate $\tau_0$ as $(\hbar/k_B T)(T_c/T)^2$ where the constant $T_c$
depends on the interaction constant between TLS and thermal
phonons.

 As for the most glasses $T_c \sim 20$ K we get at $T=0.1$ K
$\tau_0 \approx 0.3 \cdot 10^{-7}$ s. Thus at $\omega /2\pi =50$ kHz \
$\omega \tau_0 \approx 10^{-2}$. We assume the deformational potential
$\gamma_{ik} \approx 1$ eV and express the amplitude of the
deformation tensor $u_{ik}^{(0)}$. The relation between the acoustic intensity
$W$ and the deformation tensor can be expressed as
$u_{ik}^{(0)} \approx W 
/(\rho s^3)$ (where  $s$ is the sound
velocity, while $\rho$ is the density of the glass).

Demanding that the nonlinear correction should be about
10\% of the absorption we get
\begin{eqnarray}
W \ \left(\text{W}/\text{cm}^2\right) &= &0.1\,(\omega 
\tau_0)^{1/2}\, (\rho s^3 /c_d)\,\left(k_BT/\gamma \right)^2 
\nonumber \\ &&
\approx 0.03 \left[ T \, (\text{K})/0.1\right]^{1/2}
\end{eqnarray}
for $\rho =5$ g/cm$^3$, $s = 3\cdot 10^5$ cm/s.

In the case of metallic glasses we get $\tau_0 \approx
\hbar/\chi T$ with  $\chi \approx 0.01$. This gives a total
temperature dependence of $T^{-2}$ compared to $T^{-1/2}$ in the
dielectric case. The characteristic interlevel spacing in the case of
metallic glasses is $\delta = (d \, k_BT \, \omega \tau_0)^{1/2} \ll k_BT$.

Using the same values for $\gamma$, $\rho$ and $s$ for the case of
metallic glasses we get
$$ W \ (\text{W/cm}^2) \approx \rho s^3 \omega \tau_0 \left(0.1 
k_BT/\gamma \right)^2 \approx 0.01 \left[ 
T\, (\text{K})/0.1 \right]^2.$$
To conclude, we have estimated nonlinear contributions to
low-temperature absorption of acoustic waves and microwave in
dielectric and metallic glasses. It is shown that at low enough
frequencies, $\omega \tau_0 (T) \ll 1$, they are  anomalously large
and can be detected experimentally. In the high frequency regime,
$\omega \tau_0 (T) \ge 1$ the nonlinear contributions are proportional
to the intensity and do not exhibit anomalous frequency and
temperature dependencies.

\widetext
\end{document}